\documentclass[aps,pre,showpacs,showkeys,reprint,floatfix,superscriptaddress,eqsecnum,longbibliography]{revtex4-1}

\usepackage{amssymb}
\usepackage{amsmath}
\usepackage{graphicx}
\usepackage{amsbsy}
\usepackage{color}
\usepackage{bm}
\usepackage{epic}
\usepackage[normalem]{ulem}

\newcommand{\bq}{\begin{eqnarray}}
\newcommand{\eq}{\end{eqnarray}}
\newcommand{\bqn}{\begin{eqnarray*}}
\newcommand{\eqn}{\end{eqnarray*}}
\newcommand{\rr}{\mathbf{r}}

\newcommand\eff{\text{eff}}
\newcommand\HS{\text{HS}}

\begin{document}
\title{The Square-Shoulder-Asakura-Oosawa model}

\author{Riccardo Fantoni}
\email{rfantoni@ts.infn.it}
\affiliation{Universit\`a di Trieste, Dipartimento di Fisica, strada
  Costiera 11, 34151 Grignano (Trieste), Italy}

\date{\today}

\begin{abstract}
A new model for a colloidal size-asymmetric binary mixture is
proposed: The Square-Shoulder-Asakura-Oosawa. This belongs to the
larger class of non-additive hard-spheres models and has the property 
that its effective pair formulation is exact whenever the solvent
particle fits inside the interstitial region of three touching solute
particles. Therefore one can study its properties from the equivalent
one-component effective problem. Some remarks on the phase diagram of
this new model are also addressed. 
\end{abstract}



\maketitle
\section{Introduction}
\label{sec:introduction}

Hard-sphere mixtures, additive or non-additive, sticky or not, etc.,
has very rich phase diagrams showing all three phases of matter: gas, 
liquid, and solid as well as mixed or demixed and percolating or
glass. In some soft-matter laboratories
\cite{Anthony2006,Lynch2008,Russel2015}, experimentalist are 
engineering always new kinds of (Boltzmann) particles and materials
which sometimes show phase diagrams akin to the ones of
hard-spheres. It is then very important to be able to predict with
great accuracy the theoretical critical phenomena of
hard-spheres. Whereas the properties of additive (sticky-)hard-spheres,
non-additive (sticky-)hard-spheres has been carefully studied in the
past. The same attention has not been given to non-additive
square-shoulder-spheres. The one-component Square-Shoulder (SS) fluid
model has been used for the first time by Hemmer and Stell
\cite{Hemmer1970,Stell1972}. It may lead to an isostructural
solid-solid transition \cite{Bolhuis1994}, to a fluid-solid coexisting
line with a maximum melting temperature \cite{Young1977}, to unusual
phase behaviors \cite{Rascon1997,Mederos1998,Buldyrev2009} as the
reentrance of a hexatic phase in two dimensions
\cite{Prestipino2011,Prestipino2012}, and to a 
rich variety of (self-organized) ordered structures
\cite{Ziherl2001,Malescio2003,Pauschenwein2008,Fornleitner2010}. It
has been used to describe the behavior of metallic glasses 
\cite{Silbert1976,Young1977}, micellar \cite{Osterman2007} or granular
\cite{Duran1999} systems, colloidal suspensions
\cite{Lowen1993,Louis2002}, primitive models of silica
\cite{Horbach2008}, aqueous solutions of electrolytes
\cite{Galindo1999}, and water \cite{Jagla1999,Barraz2009}. The SS
model is the simplest of the class of core-softened potentials models
for fluids \cite{Rovigatti2015} that can be used. 

Recently it has been shown that augmenting the purely steric repulsion
of the Asakura-Oosawa \cite{Asakura1954} model with a {\sl soft}
repulsion shell gives rise to temperature dependent interactions which
in turn give rise to more realistic effective attractions for
cosolute-macromolecule and specifically protective osmolytes systems 
\cite{Sapir2014,Sapir2015}.

In the present work, following closely the theoretical framework of
Ref. \cite{Fantoni2015}, we will study a highly size asymmetric binary
mixture of Asakura-Oosawa where the unlike species pair-inetraction
has a square-shoulder character.

\section{Discussion}
\label{sec:discussion}

An important problem in chemical physics is that of understanding how
the behavior of the solute is influenced by the presence of the
solvent. When there is a clear distinction between which are the
solvent particles it is possible to describe the mixture as a binary
one. Imagine for example that the solvent particles are the small
``s'' ones and the solute particles are the large ``l'' ones, then for
a statistical physics description of the mixture we need to know the
potential energy of interaction between the various particles,
$U(\rr_1^{(s)},\rr_2^{(s)},\ldots,\rr_1^{(l)},\rr_2^{(l)},\ldots)$
where $\{\rr_i^{(s)}\}$ are the coordinates of the small particles and
$\{\rr_i^{(l)}\}$  the ones of the large particles. It is always
possible to write $U=U_{ss}(\rr_1^{(s)},\rr_2^{(s)},\ldots)+
U_{ll}(\rr_1^{(l)},\rr_2^{(l)},\ldots)+
U_{sl}(\rr_1^{(s)},\rr_2^{(s)},\ldots,\rr_1^{(l)},\rr_2^{(l)},\ldots)$, 
and, neglecting three bodies interactions (i.e. assuming
the particles are non-deformable, non-polarizable, $\ldots$), we can
say that $U_{ss}$ contains all pair interactions between two small
particles, $U_{ll}$ contains all pair interactions between two large
particles, and $U_{sl}$ contains all pair interactions between a small
and a large particle. Then, the influence of the solvent on the
behavior of the solute has to be due to $U_{sl}$ and $U_{ss}$. Clearly
the problem simplifies when we can assume $U_{ss}=0$ and under certain
conditions \cite{Dijkstra1999b} it can even be rewritten exactly in
terms of an {\sl effective} one-component one for only the large particle
with a potential energy $U_{ll}^\eff(\rr_1^{(l)},\rr_2^{(l)},\ldots)$
with only pairwise interactions.     

In a fluid binary mixture of small and large non overlapping hard bodies
the small particles may induce a {\sl depletion} entropic attraction
between the big particles \cite{Asakura1954,deHaro2014,Binder2014}
when two of these are closer than the dimensions of the small bodies
since in this case no small particle fit in the space between the two
large particles but there will still be an osmotic pressure due to the
small particles surrounding the two big particles pushing them
together. 

This depletion force serves an important stabilizing role in many
biological and technological processes. Specifically, many osmolytes and
polymeric crowders that are excluded from protein surfaces stabilize
the more compact folded state.\cite{Sapir2014,Sapir2015} 

Following the framework of Ref. \cite{Fantoni2015}, in order to
understand theoretically this phenomenon we will introduce 
the following model size-asymmetric binary mixtures: 
A non-additive hard-sphere binary mixture
\cite{Fantoni2011,Fantoni2013,Fantoni2014,Fantoni2015} with the
solvent particles non interacting among themselves, $\sigma_{ss}=0$,
the solute particles interacting as hard-spheres of diameter
$\sigma_{ll}=\sigma_l$, and a square-shoulder interaction between the
solvent particles and the solute particles where $\sigma_s=q\sigma_l$
is the diameter of the small spheres {\sl as seen by} the large
ones. The square-shoulder interaction occurs in a spherical shell
of diameter between $\sigma_{sl}=\sigma_l(1+q)/2$ and
$\sigma_{sl}(1+\lambda_{sl})$, the small and large particle are
otherwise interacting as hard-spheres at distances smaller than
$\sigma_{sl}$. Our size parameter $q$ here plays the role of the usual
non-additivity parameter
$\Delta=\sigma_{sl}/\sigma_{sl}^\text{add}-1=q$ with
$\sigma_{sl}^\text{add}=(\sigma_{ss}+\sigma_{ll})/2$. 

This model being non-additive (with positive non-additivity) does not
admit an analytical solution for the Percus-Yevick (PY) closure of the
Ornstein-Zernike equations but under certain geometrical condition it
admits an exact effective pair formulation \cite{Fantoni2015}. 

We called this model the Square-Shoulder-Asakura-Oosawa (SSAO) model 
\cite{Asakura1954,deHaro2014,Binder2014}. Of course, the more general
formulation of the model is when one has $\sigma_{ss}/\sigma_s$
different from 0, but in this more complicated case of an interacting
solvent we would not be able to solve exactly analitically for the
effective one-component problem \cite{Fantoni2015}. 

Usually when talking about the AO model one refers to a
colloid-polymer mixture where the depletant are linear homopolymers in
a good solvent of radius of gyration $\sigma_s/2$, which, after tracing
out the monomers degrees of freedom and replacing each chain with a
particle coinciding with its center of mass \cite{Menichetti2015}, can
be considered, for $\eta_s\lesssim 1$, to a first level of
approximation, as non interacting among themselves but unable to
penetrate a sphere of diameter $\sigma_s+\sigma_l$ around each
colloidal particles. The colloidal particles are treated as hard
spheres of diameter $\sigma_l$. In this work we will rather talk
always about a solvent-solute mixture.  
 
It is interesting to observe that when the temperature is set to
infinity the SSAO model reduces to the usual AO one with a depletion
range between $\sigma_l$ and $2\sigma_{sl}$, whereas when it
is set to zero it reduces to an AO model with a larger depletion range
extending from $\sigma_l$ to $2\sigma_{sl}(1+\lambda_{sl})$. 

To the best of our knowledge no numerical experiment has ever been
tried on the full SSAO binary mixture.

We will now first discuss the derivation of the effective
one-component problem of the SSAO model.

%
\section{The SSAO model} 
\label{sec:SSAO}

Our SSAO binary mixture model is defined as follows 
\bq \label{mI1}
U_{ss}&=&\sum_{i<j}\varphi_{ss}(|\rr_i^{(s)}-\rr_j^{(s)}|),\\
U_{ll}&=&\sum_{i<j}\varphi_{ll}(|\rr_i^{(l)}-\rr_j^{(l)}|),\\
U_{sl}&=&\sum_{i,j}\varphi_{sl}(|\rr_i^{(s)}-\rr_j^{(l)}|),\\
f_{ss}(r)&=&e^{-\beta\varphi_{ss}(r)}-1=0,\\
f_{ll}(r)&=&e^{-\beta\varphi_{ll}(r)}-1=-\theta(\sigma_{ll}-r),\\ 
\label{mI6}
f_{sl}(r)&=&e^{-\beta\varphi_{sl}(r)}-1=-\theta(\sigma_{sl}-r)+f_0(r),
\eq
where $\beta=1/k_BT$ with $T$ the absolute temperature,
$\varphi_{ss},\varphi_{ll},\varphi_{sl}$ are the bare 
solvent-solvent, solute-solute, and solvent-solute pair-potentials
respectively, $f_{ss}, f_{ll}, f_{sl}$ the corresponding Mayer
functions, $\sigma_{sl}=\sigma_l(1+q)/2$, $\theta$ is the Heaviside 
step function, and the square-shoulder is
\bq
f_0(r)=\left\{\begin{array}{ll}
0&r<\sigma_{sl}\\
e^{-\beta\epsilon_{sl}}-1& \sigma_{sl}<r<\sigma_{sl}(1+\lambda_{sl})\\
0&r>\sigma_{sl}(1+\lambda_{sl})
\end{array}
\right.,
\eq
where $\epsilon_{sl}>0$ is a positive constant. We can then introduce
a reduced temperature as $T^*=1/\beta\epsilon_{sl}$. 

Model SSAO does not admit a PY analytic solution but admits an exact
effective one-component description for
$q<q_0=2/\sqrt{3}-1=0.15470\ldots$, when a solvent can fit into the  
inner volume created by three solutes at contact (or $q<1$ in one
spatial dimension \cite{Brader2002}), so that a solvent particle
cannot overlap simultaneously with more than two (nonoverlapping)
solutes at contact. Following the
derivation of Dijkstra et al. \cite{Dijkstra1999b}, we describe the
mixture in a mixed canonical (for the solutes) and grand canonical
(for the solvent) ensemble, which they call semi-grand-canonical
$(z_s,N_l,V,T)$ ensemble. It is then easy to show \cite{Fantoni2015}
that in this case, after integrating out the degrees of freedom of the
solvent, the effective potential $\beta v_{ll}(r)$ is 
\bq
\beta v_{ll}(r)=\beta\varphi_{ll}(r)-z_s\int d\rr_s\,f_{sl}(r_s)
f_{sl}(|\rr_s-\rr|),
\eq
which upon using Eq. (\ref{mI6}) gives
\bq \label{ep}
\beta v_{ll}(r)=\left\{\begin{array}{ll}
+\infty&r<\sigma_l\\
v_{0}(r)&\sigma_l<r<2\sigma_{sl}(1+\lambda_{sl})\\
0&r>2\sigma_{sl}(1+\lambda_{sl}) 
\end{array}
\right..
\eq
with
\begin{widetext}  
\bq \nonumber
&&-v_{0}(r)/z_s=\\ \nonumber
&&2{\cal C}(r/2,\sigma_{sl})-\\ \nonumber
&&2\left(e^{-\beta\epsilon_{sl}}-1\right)
[{\cal C}(r_<,\sigma_{sl})+{\cal C}(r_>,\sigma_{sl}(1+\lambda_{sl}))-
2{\cal C}(r/2,\sigma_{sl})]+\\ \label{epswao}
&&2\left(e^{-\beta\epsilon_{sl}}-1\right)^2\{
{\cal C}(r/2,\sigma_{sl}(1+\lambda_{sl}))-
[{\cal C}(r_<,\sigma_{sl})+{\cal C}(r_>,\sigma_{sl}(1+\lambda_{sl}))-
{\cal C}(r/2,\sigma_{sl})]\},
\eq
where we denoted with ${\cal C}(R,\sigma)$ the volume of a spherical
cap of height $\sigma-R$ in a sphere of radius $\sigma$, i.e.
\bq
{\cal C}(R,\sigma)=\frac{2\pi\sigma^3}{3}
\left[1-\frac{3}{2}\frac{R}{\sigma}+
\frac{1}{2}\left(\frac{R}{\sigma}\right)^3\right]\theta(\sigma-R),
\eq
and $r_<+r_>=r$ with
\bq
r_<&=&\frac{r^2+\sigma_{sl}^2-\sigma_{sl}^2(1+\lambda_{sl})^2}{2r},\\
r_>&=&\frac{r^2+\sigma_{sl}^2(1+\lambda_{sl})^2-\sigma_{sl}^2}{2r}.
\eq 
\end{widetext}
In this case, from Dijkstra et al. \cite{Dijkstra1999b} derivation,
one finds that the exact cancellation of all Meyer diagrams higher
than the two body one, occurs when \cite{Fantoni2015}
\bq \nonumber
\frac{\sigma_{sl}(1+\lambda_{sl})-\sigma_l/2}{\sigma_l/2}&=&
q+\lambda_{sl}+q\lambda_{sl}<q_0\\
&=&\frac{2}{\sqrt{3}}-1.
\eq
It is easy to show \cite{Fantoni2015} that in the range
$\sigma_l<r<2\sigma_{sl}(1+\lambda_{sl})$ one finds the result of
Eq. (\ref{ep}) for the effective potential. 

This is formally the same effective potential found in
model SWAO of Ref. \cite{Fantoni2015} where we change
$\epsilon_{sl}\to-\epsilon_{sl}$. In the present case the sticky limit
procedure would give the same AO model.

\begin{figure}[htbp]
\begin{center}
\includegraphics[width=8cm]{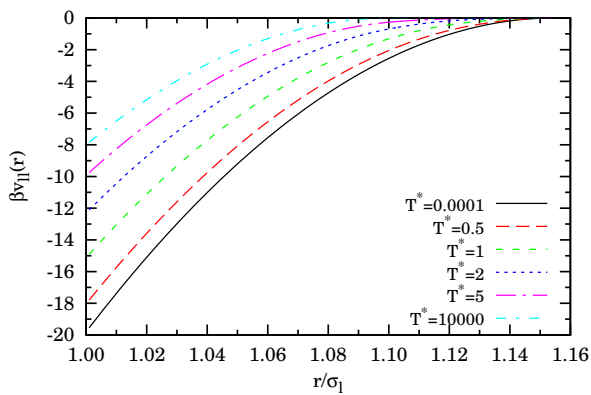}
\end{center}  
\caption{(Color online) Effective potential for the SSAO mixture at
  various reduced temperatures and $q=0.1, \lambda_{sl}=(q_0-q)/(1+q),
  \eta_s^{(r)}=1/2$.}
\label{fig:poteff}
\end{figure}

In Fig. \ref{fig:poteff} we show the effective potential for the SSAO
binary mixture at various temperatures.
From the figure we see how at infinite temperature the SSAO reduces to 
the usual AO model with a range $2\sigma_{sl}$ whereas at zero
temperature it reduces to an AO model with a wider range extending to
$2\sigma_{sl}(1+\lambda_{sl})$. At intermediate temperatures the
effective potential lies continuously between the two extreme cases.
Thus, the soft shoulder repulsion enhances the depletion
mechanism and the solute stabilization.

Here $z_s$ is the solvent fugacity. We can introduce a solvent
reservoir packing fraction 
$\eta_s^{(r)}=\pi z_s\sigma_s^3/6=\eta_s e^{\beta\mu_s^\text{ex}}$,
with $\mu_s^\text{ex}$ the excess (over the ideal gas) solvent
chemical potential. The solvent reservoir is at the same temperature
of the solute. $\eta_s^{(r)}$ is the packing fraction of the reservoir
made of noninteracting solvent particles. The relationship between
$\eta_s^{(r)}$ and $\eta_s$ can be found calculating the average
number of small solvent particles (see appendix C in
Ref. \cite{Fantoni2015}). Up to second order in $\eta_l$ one finds
\bq \nonumber
\eta_s&\approx&\eta_s^{(r)}\left.\Bigg[1+\eta_l(1+q)^3
(\lambda_{sl}(3+\lambda_{sl}(3+\lambda_{sl}))e^{-\frac{1}{T^*}}\right.\\ \nonumber
&&\left.-(1+\lambda_{sl})^3)-\frac{12\eta_l^2q^3}{\sigma_l^3}\times\right.\\ 
\label{resden}
&&\left.\int_{\sigma_l}^{\sigma_l(1+q)(1+\lambda_{sl})}
dr\,r^2(v_0/\eta_s^{(r)})e^{-v_0(r)}\right]. 
\eq

\section{The Noro and Frenkel criterion} 
\label{sec:NF}

Noro and Frenkel \cite{Noro2000} argued that the reduced second virial
coefficient $B_2/B_2^\text{HS}$, rather than the range and the
strength of the attractive interactions, could be the most convenient
quantity to estimate the location of the gas-liquid critical point for
many different colloidal suspensions. Their criticality criterion for
particles with variable range attractions, complemented by
the simulation value of the critical temperature obtained in
Ref. \cite{Miller2004} for the SHS model, yields $B_2/B_2^\text{HS}\approx
-1.21$.

Applying Noro and Frenkel criticality criterion for particles with
variable range attractions \cite{Noro2000}, complemented by
the simulation value of the critical temperature obtained in
Ref. \cite{Miller2004} for the Sticky-Hard-Sphere model, to our
effective one-component problem, we are led to conclude that
criticality requires $B_2^\text{eff}/B_2^\HS=-1.21$ where
$B_2^\text{eff}$ is the second virial coefficient of our effective
solute-solute problem 
\bq
B_2^\text{eff}=\frac{2\pi}{3}\left\{\sigma_l^3-3\int_{\sigma_l}^\infty dr\,r^2
\left[e^{-\beta v_{ll}(r)}-1\right]\right\},
\eq
and $B_2^\HS=2\pi\sigma_l^3/3$ is the virial coefficient for HS of
diameter $\sigma_l$. Note that $B_2^\text{eff}$ can only be calculated
numerically. 

\begin{figure}[htbp]
\begin{center}
\includegraphics[width=8cm]{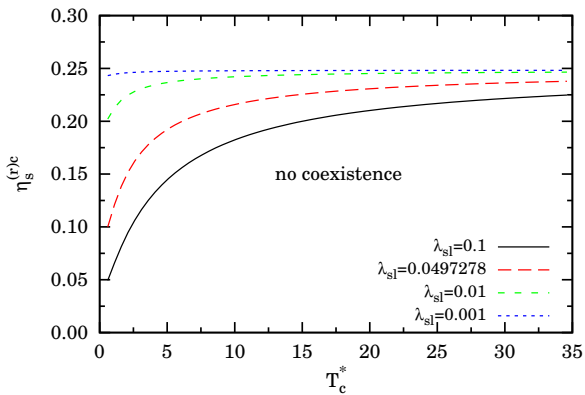}
\end{center}  
\caption{(Color online) Phase diagram stemming from the Noro and
  Frenkel empirical criterion in the $(T^*,\eta_s^{(r)})$ plane, 
  for a highly asymmetric, $q=0.1$, SSAO for four values of
  $\lambda_{sl}$.} 
\label{fig:pd}
\end{figure}

In Fig. \ref{fig:pd} we show the coexistence curves for the phase
diagram stemming from the Noro-Frenkel empirical criterion in the
$(T^*,\eta_s^{(r)})$ plane, for the SSAO model with $q=0.1$ for four 
values of $\lambda_{sl}=0.001, 0.01, (q_0-q)/(1+q), 0.1$. For the last
$\lambda_{sl}=0.1$ case the effective potential of Eq. (\ref{ep}) is
only an approximation.  
From the figure we see how even a small shoulder produces dramatic
effects in the phase diagram, widening the fluid-fluid coexistence
region. 
 
\section{Perturbation theory for the SSAO model} 
\label{sec:PTII}

From the previous section we understood that the {\sl hidden} or
metastable fluid-fluid phase separation observed by Dijkstra et
al. \cite{Dijkstra1999b} in their study of the AO model could be
enlarged by adding a soft unlike repulsion as in our SSAO model. Now
we want to quantify this more precisely through first order
perturbation theory \cite{Gast1983}. Taking the HS as reference system
we can write the Helmholtz free energy per particle, $a=A/N$, as
follows 
\bq \nonumber
\beta a_{ll}&=&\beta a_\HS+\\ \label{ptaII}
&&12\eta_l\int_{\sigma_l}^{\sigma_l(1+q)(1+\lambda_{sl})}\beta
v_{ll}(r)g_\HS(r)r^2\,dr,
\eq
where $\beta
a_\HS=(4\eta_l-3\eta_l^2)/(1-\eta_l)^2+\ln(\eta_l)+\mbox{constants}$ 
is the Carnahan-Starling \cite{Carnahan69} expression for HS, $\beta
v_{ll}$ the effective pair-potential of the SSAO model of
Eq. (\ref{ep}), and $g_\HS$ is the HS radial distribution function in
the PY approximation \cite{Smith1970}, which in the interval
$1<r<(1+q)(1+\lambda_{sl})<2$ can be written as follows  
\bq
rg_\HS(r)=\sum_{i=0}^2\lim_{t\to t_i}(t-t_i)t\frac{L(t)}{S(t)}e^{t(r-1)},
\eq  
where we are measuring lengths in units of $\sigma_l$,
\bq \nonumber
S(t)&=&(1-\eta_l)^2t^3+6\eta_l(1-\eta_l)t^2+18\eta_l^2t-\\
&&12\eta_l(1+2\eta_l),\\
L(t)&=&(1+\eta_l/2)t+1+2\eta_l,
\eq
and $t_i (i=0,1,2)$ are the zeros of $S(t)$. The first order Helmholtz
free energy of Eq. (\ref{ptaII}) can thus be calculated analytically.

The compressibility factor $Z=\beta p/\rho$ and chemical potential
$\mu$ are then found through
\bq
Z_l&=&\eta_l\left.\frac{\partial\beta
  a_{ll}}{\partial\eta_l}\right|_{\eta_s^{(r)}},\\
\beta\mu_l&=&Z_l+\beta a_{ll}.
\eq
The critical point $(\eta_s^{(r)c},\eta_l^c)$ is determined by
numerically solving the following system of equations 
\bq 
\left.\frac{\partial Z_l\eta_l}
{\partial\eta_l}\right|_{\eta_s^{(r)c},\eta_l^c}&=&0,\\
\left.\frac{\partial^2 Z_l\eta_l}
{\partial\eta_l^2}\right|_{\eta_s^{(r)c},\eta_l^c}&=&0.
\eq

\begin{figure}[htbp]
\begin{center}
\includegraphics[width=8cm]{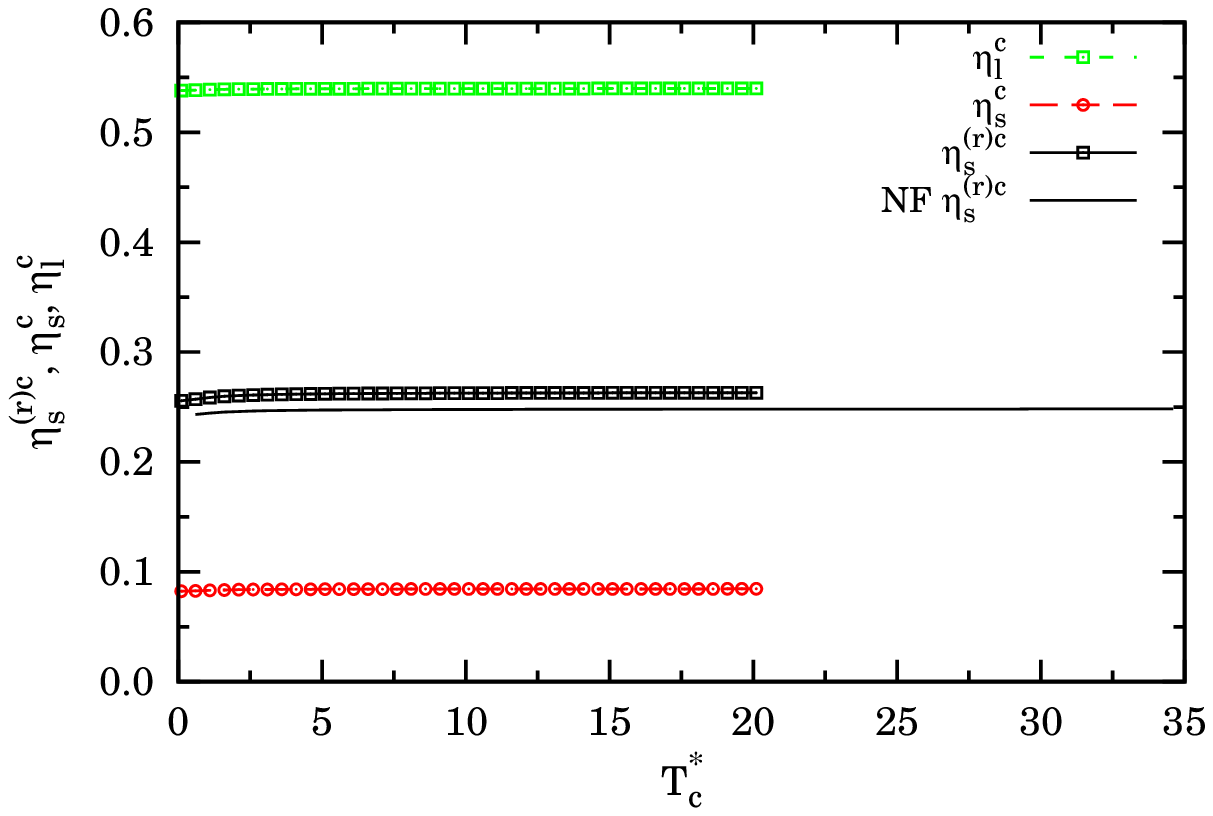}\\
\includegraphics[width=8cm]{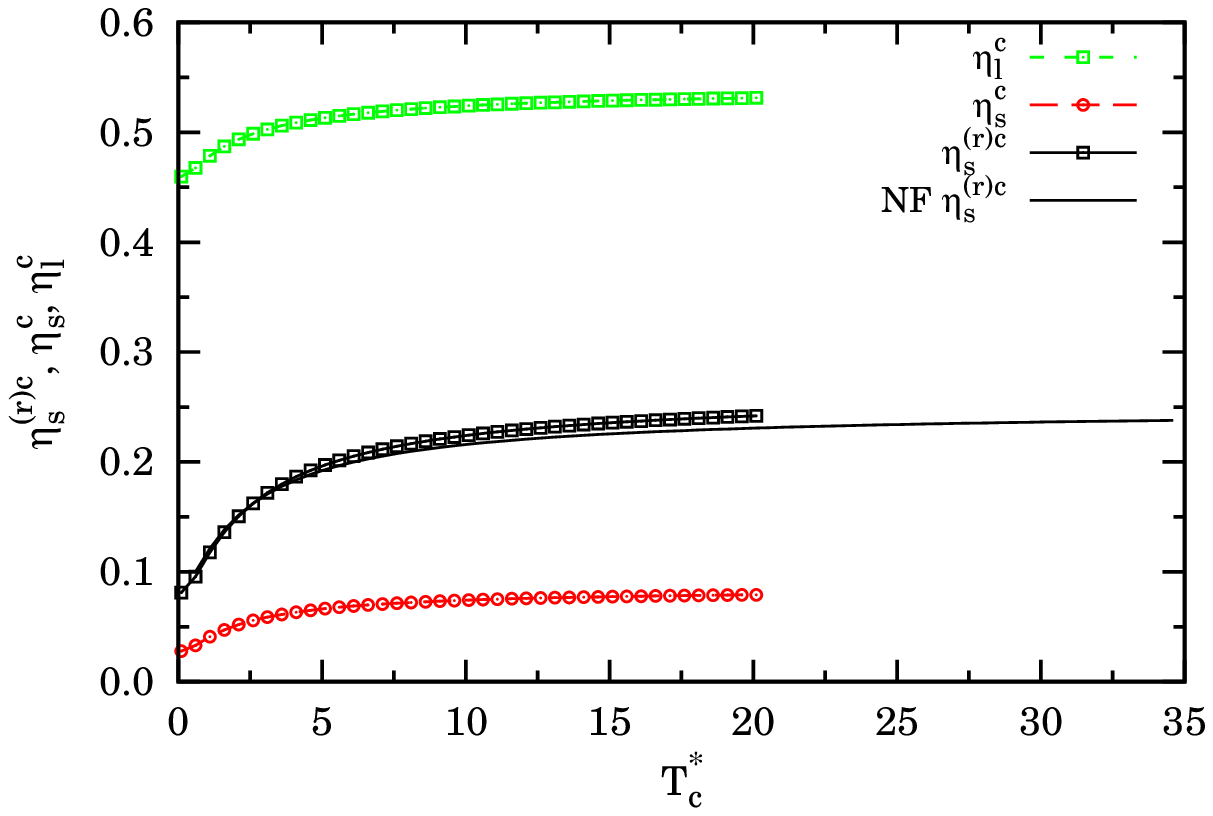}
\end{center}  
\caption{(Color online) Critical point for the fluid-fluid coexistence
  in SSAO for $q=0.1$ and $\lambda_{sl}=0.001$ (top panel) and
  $\lambda_{sl}=(q_0-q)/(1+q)$ (bottom panel) as a function of
  $T^*_c$. The lines with symbols are obtained from thermodynamic
  perturbation theory, whereas the solid line corresponds to the Noro
  and Frenkel criterion of Fig. \ref{fig:pd}. Eq. (\ref{resden}) is
  used for the conversion between the reservoir density and the
  solvent density.} 
\label{fig:cp}
\end{figure}

In Fig. \ref{fig:cp} we show the critical point
$(\eta_s^{(r)c},\eta_l^c)$ for the fluid-fluid coexistence of the SSAO
model for $q=0.1$ and $\lambda_{sl}=0.001$, near to the AO,
and $\lambda_{sl}=(q_0-q)/(1+q)$, the full SSAO, as a function of the
reduced temperature, $T^*$.
The figure confirms the scenario predicted in the previous section
from the Noro-Frenkel criterion but gives additional information on
the critical solvent and solute packing fractions, $\eta_s^c$ and
$\eta_l^c$ respectively.
Of course we expect a breakdown of the perturbation theory treatment
as soon as the depletion mechanism becomes too strong. Also as soon as
$q>q_0$ we are neglecting three-body (and higher) terms.  

\section{Conclusions} 
\label{sec:conclusions}

We studied a new colloidal strongly asymmetric binary mixture of
(small) solvent and (large) solute particles, where unlike particles
interact through the repulsive Square-Shoulder (SS) pair-potential,
that we called the Square-Shoulder-Asakura-Oosawa (SSAO)
model. Whenever the solvent particle fits inside the interstitial
region of three touching solute particles we were able to derive
exactly analytically an effective solute-solute pair-potential and
discussed the corresponding phase behaviors, as obtained from the
resulting effective one-component system. 

We found that the mere presence of the soft repulsion in the spherical 
shell of the SS unlike pair-interaction augments the depletion
mechanism typical of the underlying Asakura-Oosawa (AO)
mixture. Applying the Noro and Frenkel criterion we saw that this, in
turn, enlarges (and may stabilize) the metastable fluid-fluid phase
coexistence region typical of the strongly asymmetric AO model at
large reservoir packing fractions. A first order thermodynamic
perturbation theory nicely confirms the scenario depicted by such
criterion.

This phenomenon can be relevant in the experimental study of colloidal
suspensions undergoing a fluid-fluid phase transition in the
laboratory. Whenever the mathematical mixture just described
represents a good model for a real mixture it should be expected that
the main effect of the presence of the additional repulsive shell in
the unlike species pair-interaction is to increase the depletion
mechanism and in turn to enlarge the phase coexistence region of the
phase diagram.


\begin{acknowledgments}
We acknowledge fruitful discussions with Andr\'es Santos.
\end{acknowledgments} 
\bibliography{ssao}
\end{document}